\documentclass{cta-author}
\usepackage[T1]{fontenc} 
\usepackage{fancyref}
\usepackage{amsmath,eqnarray}
\usepackage{amsthm,amsfonts}
\usepackage{commath}
\usepackage{enumitem}
\setlist[enumerate]{leftmargin=*,
    labelindent=.5\parindent,
    labelsep=.5\parindent}
\setlist[itemize]{leftmargin=*,
    labelindent=.5\parindent,
    labelsep=.5\parindent}
\usepackage[cmintegrals]{newtxmath}
\usepackage{bm} 
\usepackage{amssymb}
\usepackage{graphicx}
\usepackage{subcaption}
\usepackage{natbib}
\usepackage{hyperref}
\hypersetup{
    colorlinks=true,
    linkcolor=blue,
    filecolor=blue,      
    urlcolor=blue,
    citecolor=blue
}

\def\subsectionautorefname{Subsection}
\theoremstyle{definition}
\newtheorem{lem}{Lemma}
\newtheorem{rem}{Remark}
\begin{document}
\supertitle{Research Article}
\title{Ergodic Spectral Efficiency of Massive MIMO with Correlated Rician Channel and MRC Detection based on LS and MMSE Channel Estimation}
\author{\au{Mohammad Hadi Sadraei $^1$} \au{Mohammad Sadegh Fazel $^2$}  \au{Ali Mohamad Doost-Hoseini $^3$}}
\address{\add{1}{Department of Electrical and Computer Engineering, Isfahan University of Technology, Isfahan 84156 83111, Iran\email{mh.sadraei@ec.iut.ac.ir}}
\add{2}{Department of Electrical and Computer Engineering, Isfahan University of Technology, Isfahan 84156 83111, Iran\email{fazel@iut.ac.ir}}
\add{3}{Department of Electrical and Computer Engineering, Isfahan University of Technology, Isfahan 84156 83111, Iran\email{alimdh@iut.ac.ir}}}
\begin{abstract}
    In this paper, we study the spectral efficiency (SE) of a multi-cell massive multiple-input multiple-output (MIMO) system with a spatially correlated Rician channel. The correlation between least squares (LS) estimator and its error complicates SE analysis,
    since signal and interference components become cross-correlated, too.
    Minimum mean square error (MMSE) estimators do not suffer from this burden.
    In some previous works, a proper part of the signal is referred to interference, which makes them cross-uncorrelated, and leads to a SE lower bound.
    In our modified approach, we extract and refer the cross-correlated part of interference to the signal to attain this objective. Here, we use this approach for calculating the instantaneous SE of maximum ratio combining (MRC)
    detector under LS and MMSE estimation methods. We further derive closed-form approximations of their ergodic
    SE. This approach is also applicable to other linear channel estimators or data detectors. Numerical results show that achievable
    SE surpasses that of the previous approach. Moreover, they show that our approximation is close enough to Monte Carlo simulation results,
    especially at the high number of the base station (BS) antennas.
    
    \emph{Index terms}: Massive MIMO, Uplink, Spectral Efficiency, LS Channel Estimation, MMSE Channel Estimation
\end{abstract}
\maketitle
\graphicspath{{Revision2-Figures/}}
\section{Introduction}\label{intro}
Higher SE with a reliable transmission is an obvious requirement in
wireless communication systems, since the available spectrum is
saturated  \cite{5gbe}.
Massive MIMO system is one of the solutions for
improving SE \cite{7581208}.
In this system, the BS in each cell is equipped with a large number of
antennas compared to active users \cite{maminext}.
In perfect channel~state~information~(CSI) case of uncorrelated Rayleigh channel, cross-interference, and uplink
thermal noise effects vanish by a high increase in the number of BS antennas, as a result of random matrix properties \cite{noncoop,intromassive}.
Moreover, linear
processing at BS can provide an achievable sum-rate close to optimal
non-linear solutions like maximum~likelihood~(ML) detection in
uplink and dirty~paper~coding~(DPC) in downlink \cite{tenmyths}.

When there is a strong line of sight~(LOS) path and a large number of independent non-LOS paths between users and BS, the transmission channel exhibits a Rician model.
In the absence of the LOS component, it follows a Rayleigh model, which is a particular case of Rician \cite{Tse05fundamentalsof}.
Furthermore, these distributions have been used to model the channel in 5G standard \cite{ETSI5G}.
However, some measured channels~\cite{892532} do not fit into either of them.
Hence, other distributions like Nakagami \cite{NAKAGAMI_1960} is also used to model the fading channel. Luckily, this distribution can be approximated as Rician \cite{goldsmith_2005}.
Thus, system performance with Nakagami fading can be approximately analysed by using results achieved by the Rician channel.

SE of massive MIMO systems has been mostly investigated for uncorrelated Rayleigh channel in both uplink \cite{7956175,energyspectral,achieveestimate,upzf,8240588,2014.0566,8103416,2016.0538} and
downlink \cite{2016.0538,conjugate,7279083}.
Among these works, the single-cell scenario is assumed in
\cite{energyspectral,achieveestimate,8240588,2014.0566,conjugate}
and multi-cell in
\cite{7956175,upzf,8103416,2016.0538,7279083}.
SE of MRC detection is investigated for three cases: perfect CSI \cite{energyspectral};
LS \cite{7956175} and
MMSE channel estimation \cite{energyspectral,achieveestimate,7956175}.
SE of zero forcing (ZF) detector is studied for perfect CSI \cite{energyspectral}, as well as imperfect CSI (MMSE channel estimation) cases \cite{energyspectral,achieveestimate,upzf}.
SE of MMSE detection is evaluated in \cite{energyspectral} for both perfect CSI and MMSE channel estimation cases.
SE of ML detector is approximated for a multi-cell system in \cite{2014.0566} under perfect CSI assumption.
In \cite{8103416}, ZF detection is modified to have less inter-cell interference and, as a result, higher SE in the presence of MMSE channel estimation.

In \cite{2016.0538}, both uplink and downlink SE of MRC/MRT and ZF processing at BS are studied for uncorrelated Rayleigh channel with MMSE estimation.
Lower bounds for SE of maximum ratio transmission~(MRT) and ZF pre-coders are provided for MMSE channel estimation in \cite{conjugate}.
Achievable SE of MRT and ZF pre-coders are provided in \cite{7279083}, whether with or without considering downlink pilots.

Correlated Rayleigh channel is also considered for single-cell \cite{7018998} and multi-cell systems \cite{8603076}.
In a more practical case,
the effect of covariance estimation error on both uplink and downlink SE of MMSE channel estimation is studied in \cite{8603076}.
In \cite{7018998}, ZF pre-coding is modified to achieve higher SE of the perfect CSI case.

\subsection{Related Literature}
SE approximations of massive MIMO Rician channels are provided in \cite{8643884,acratefdup,impalos,ozdtranc} for the uplink including MRC \cite{acratefdup,impalos,ozdtranc} and ZF detector \cite{8643884}; and \cite{ozdtranc,appzfrice,aghaeinia,acratefdup} for the downlink
covering MRT \cite{aghaeinia,acratefdup,ozdtranc} and ZF pre-coder \cite{appzfrice}.
However, spatial correlation is considered only in \cite{impalos,ozdtranc}.
Perfect CSI is assumed in \cite{impalos} while in \cite{ozdtranc} non-ideal LS and MMSE channel estimations are taken into account.
Imperfect CSI is also considered in \cite{aghaeinia}, but for uncorrelated Rician channel.
In \cite{ozdtranc,aghaeinia}, a multi-cell system is considered, while in the other mentioned works single-cell scenario is studied.
\subsection{Contribution}
To the best of our knowledge, SE analysis for correlated Rician channel and imperfect CSI is presented only in \cite{ozdtranc}.
In this work, lower bounds are provided for the signal to interference plus noise ratio (SINR),
which are based on the mean of the effective channel (i.e., the cascade of the channel and the detector) \cite{massivemimobook}.
In this view, signal and interference decorrelate, but less of the available CSI is used.

In our work,
the SE of correlated Rician channel for imperfect CSI case is analysed with a new approach.
Here, SINR of each user is calculated based on the equivalent channel which is
the cascade of the estimated channel and the detector.
Hence, all available CSI is used.
However, signal and interference are not necessarily cross-uncorrelated, which makes SE analysis more difficult.
We overcome this difficulty by extracting
the correlated component of interference with the signal and adding it to the desired signal part.
Thus, modified desired signal and interference become cross-uncorrelated to afford
instantaneous SE calculation.
Besides, some near-optimal closed-form formulas are also derived for ergodic SE.
To sum up, the main contributions of this paper are as follows:
\begin{itemize}
    \item
    In a multi-cell correlated Rician channel, we propose to extract all data-dependent components at the detector output as a signal.
    Therefore instantaneous SE of imperfect CSI can be calculated in the form of ${\log}_2\bigl( 1+\nu \bigr)$,
    where $\nu$ is the SINR.
    \item
    An approximation is proposed for closed-form ergodic SE of the imperfect CSI case. Necessary statistics are further derived by using quadratic and quartic moments of a complex normal vector.
    These are derived by using sufficient statistics of MMSE and LS estimators.
    \item 
    We compare our proposed approximation with Monte Carlo simulation results, as well as \cite{ozdtranc} for both single-cell and multi-cell correlated Rician channels.
    Moreover, we show the superiority of our work and the closeness of the proposed approximation to simulation results.
\end{itemize}
\subsection{Outline}
The rest of the paper is organised as follows: In \autoref{model}, the system model, pilot, and data transmission processes are discussed.
We propose our SE analysis and ergodic SE approximation in \autoref{rate}.
Numerical results are provided in \autoref{sim}.
Finally, the paper is concluded in \autoref{conc}.
\subsection{Notation}
Vectors and matrices are italic boldface lower and higher cases, respectively.
Superscript ${(\cdot)}^{H}$ 
denotes complex conjugate transpose~(hermitian) of a vector or matrix.
The trace of $\bm{X}$ is shown by $\mathrm{Tr}\bigl\lbrace \bm{X} \bigr\rbrace $.
Symbols $\mathrm{E}[ \bm{x} ]$ and $\| \bm{x} \|$ denote the expected value and Frobenius norm of the vector $\bm{x}$, respectively.
Set of all complex matrices with $K\times M$ size and
vectors with $K$ elements
are denoted by ${{\mathbb{C}}^{K\times M}}$ and ${\mathbb{C}}^{K}$, respectively.
The identity matrix is indicated by $\bm{I}$.
Real and complex normal~vectors with mean vector $\bm{m}$ and  covariance matrix
$\boldsymbol{\Sigma}$ are shown by 
$\mathrm{N}\bigl( \bm{m},\boldsymbol{\Sigma} \bigr)$
and $\mathbb{C}\mathrm{N}\bigl( \bm{m},\boldsymbol{\Sigma} \bigr)$, respectively.
Finally, uniform distribution with minimum and maximum values of $a$ and $b$ is denoted by $\mathbb{U}\bigl( a,b\bigr)$.
\section{System Model}\label{model}
\begin{figure}[!t]
    \centering
    \includegraphics[width=\linewidth]{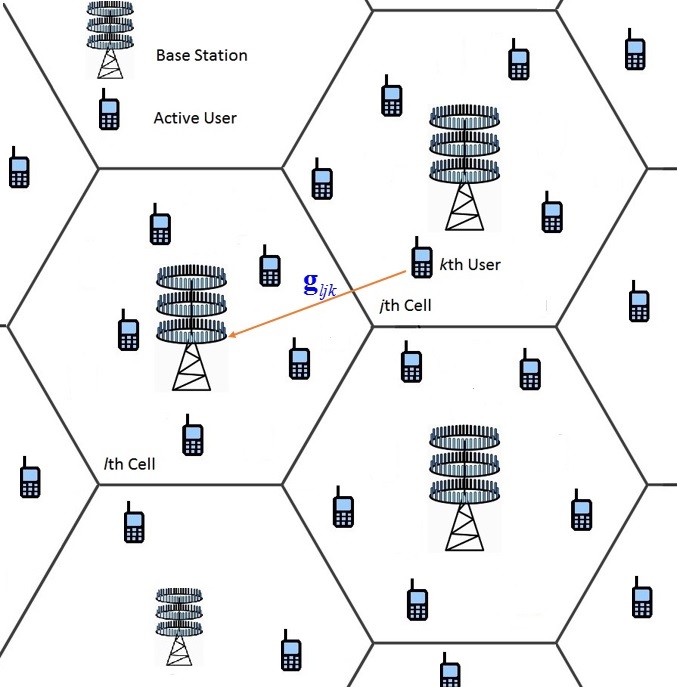}
    \caption{Multi-cell massive MIMO scenario.}
    \label{fig:system model}
\end{figure}
We consider the uplink of a system with $L$ cells (\autoref{fig:system model}), where each BS contains $M$ antennas and serves $K$ single-antenna users.
Here, orthogonal frequency division multiplexing~(OFDM) is used such that no inter-symbol-interferences (ISI) and
inter-carrier-interferences (ICI) exist.
However, some few works have considered ISI \cite{8691003,8880696} or ICI \cite{8853272} in massive MIMO systems.
It is known that SE decreases in the presence of either ISI or ICI. 
However, it is shown in \cite{8691003} that using an MRC detector reduces ISI in massive MIMO systems itself.
Also, parallel quadrature spatial modulation has been proposed in \cite{8880696} to compensate ISI.
In \cite{8853272} performance of a linear MMSE equaliser has been investigated for a large MIMO-OFDM system considering carrier frequency off-set along with some other hardware impairments.

Within each coherence block, the channel between $k$th user in $j$th cell and BS in $l$th cell
is described as vector $\bm{g}_{ljk}\in {{\mathbb{C}}^{M}}$ which has a $\mathbb{C}\mathrm{N}\bigl( \bm{m}_{ljk}, \bm{R}_{ljk} \bigr)$ distribution.
Thus, the magnitudes of elements of $\bm{g}_{ljk}$ have a Rician distribution.
Non-zero off-diagonal elements of $\bm{R}_{ljk}$ represent channel spatial correlation.
The vectors $\bm{g}_{ljk}$ are assumed to be independent for different values of
$\bigl( l,j,k\bigr)$ because users are widely distributed in each cell.
The mean vectors ($\bm{m}_{ljk}$) correspond to the LOS components and depend on the large scale fading factors (${\beta}_{ljk}^{LOS}$).
The covariance matrices ($\bm{R}_{ljk}$) are related to non-LOS paths and depend on their large scale fading multiples (${\beta}_{ljk}^{NLOS}$).

It is assumed that the location of each user is approximately fixed in each coherence block.
In a long time, the movement of each user changes the distance between the user and BS, as well as the corresponding large scale fading multiple.
Besides, the speed of variation in small scale fading coefficients depends on the velocity of the user.
Here, we assume that the small scale fading coefficients do not change within the coherence block but change between the blocks.
By increasing the velocity such that the channel does not remain static in a coherence block, its length must be reduced according to the new coherence time of the channel.
The analysis is still applicable to the new coherence block.

In this paper, imperfect CSI is assumed, and only channel statistics are perfectly available.
However, in practice, a long-time averaging is needed to provide accurate statistics.
Therefore, it is necessary to estimate these statistics, which have errors. These estimation errors degrade SE \cite{8603076}.
\begin{figure}[!t]
    \centering
\includegraphics[width=\linewidth]{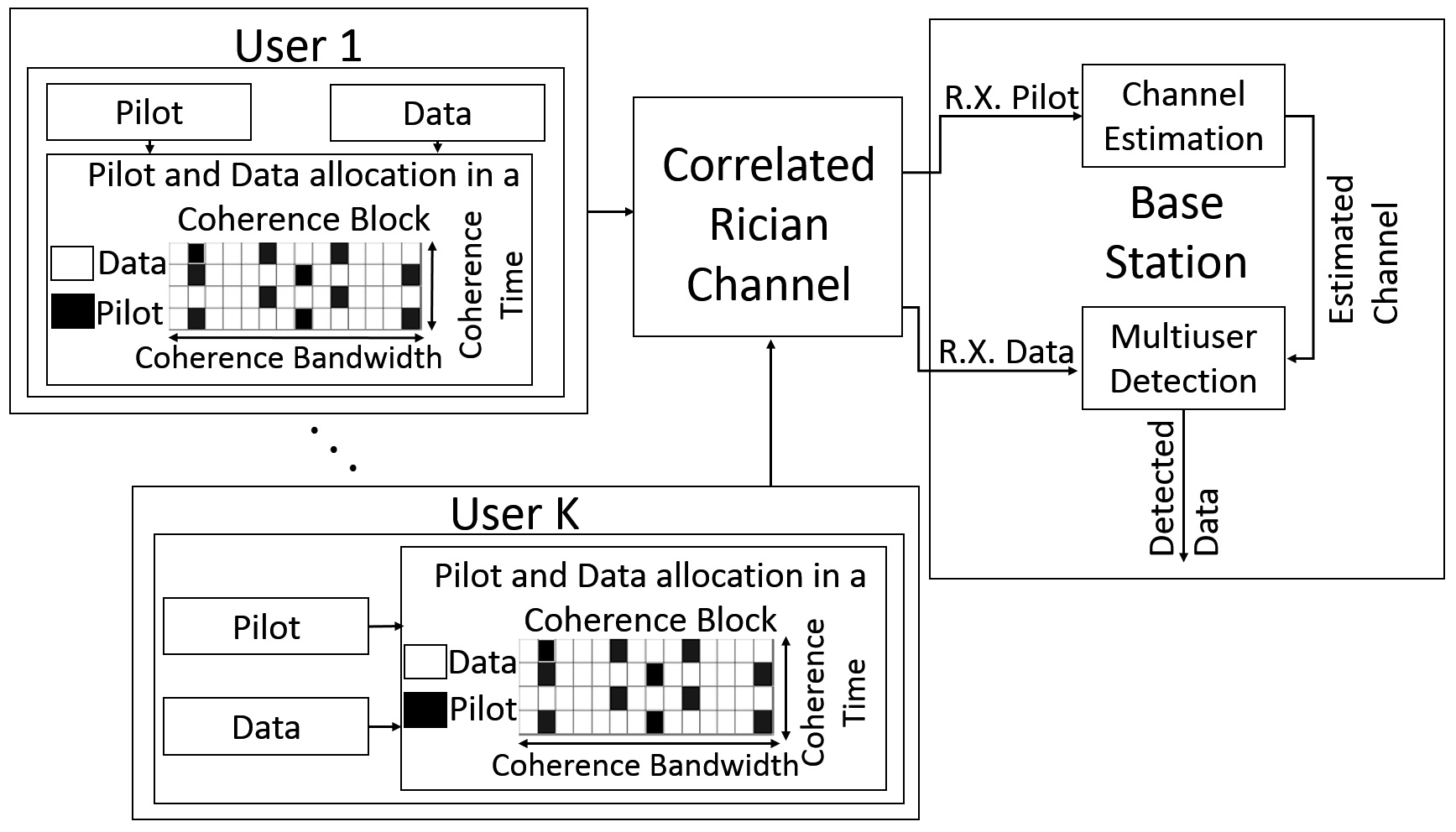}
    \caption{Block diagram of pilot and data allocation at users; channel estimation and data detection at BS over one coherence block.}
    \label{fig:blockdiagram}
\end{figure}

As seen in \autoref{fig:blockdiagram}, in each coherence block, BS estimates the channel and detects the data. Channel is estimated by processing the received pilot
sequences.
Then, data is detected by using the estimated channel.
\subsection{Channel Estimation}\label{est}
We assume that $k$th user in $l$th cell transmits the pilot sequence $\sqrt{q_{lk}}{\boldsymbol{\phi}}_{k}$ of length ${\tau}_p$, where
$q_{lk}$ is symbol power and ${\boldsymbol{\phi}}_{k}$ is such that $\|{\boldsymbol{\phi}}_{k}\|^2={\tau}_p$.
By assuming $\tau_{p}\geq K$, we can have $K$ mutually orthogonal sequences.
The pilot sequence length is limited due to the limitation in the coherence block length.
Therefore, when the number of users is high, it is impossible to assign orthogonal pilot sequences to users in all cells. Hence, some users in different cells have to send the same pilot sequence.
Here, it is assumed that users in each cell transmit orthogonal pilot sequences, but users in other cells send similar pilot sequences.
Hence, the received pilot by each BS is contaminated by transmitted pilots from users in adjacent cells.
This phenomenon is called pilot contamination.

We restrict our analysis to the $l$th cell.
For simplicity the index $l$ is dropped for channel vector between all users and BS in $l$th cell 
(${\bm{g}}_{llk}$ for all $k$), as well as its mean vector (${\bm{m}}_{llk}$) and covariance matrix (${\bm{R}}_{llk}$).
In other words, ${\bm{g}}_{llk}$, ${\bm{m}}_{llk}$, and ${\bm{R}}_{llk}$ are written as ${\bm{g}}_{k}$, ${\bm{m}}_{k}$, and ${\bm{R}}_{k}$, respectively for all $k$.
The received pilot at BS ($\bm{\Psi}\in {\mathbb{C}}^{M\times {\tau}_p}$) is
\begin{equation}\label{pilottranmission}
\bm{\Psi}=\sum\limits_{j=1}^{L}{\sum\limits_{i=1}^{K}{\sqrt{{{q}_{ji}}}{{\bm{g}}_{lji}}{\boldsymbol{\phi}}_{i}}}+{\bm{W}},
\end{equation}
where ${\bm{W}}\in {{\mathbb{C}}^{M\times {{\tau }_{p}}}}$
is the additive white Gaussian noise (AWGN) matrix 
with independent and identically distributed (i.i.d)
$\mathbb{C}\mathrm{N}\bigl( 0,\sigma^2_n \bigr)$ entries.
At BS, the matrix-vector product $\bm{\Psi}{\boldsymbol{\phi}}_{k}^{H}$ possesses sufficient statistics for estimating ${\bm{g}}_{k}$:
\begin{equation}\label{suff}
\bm{\Psi}{\boldsymbol{\phi}}_{k}^{H}=
{\tau}_{p}\sqrt{{q}_{lk}}{\bm{g}}_{k}+
{\tau}_{p}\sum\limits_{\substack{j=1 \\ j\neq l}}^{L}\sqrt{{q}_{jk}}{\bm{g}}_{ljk}+{\bm{W}}{\boldsymbol{\phi}}_{k}^{H}.
\end{equation}
The second and third terms are pilot contamination and noise effects, respectively.
The vector $\bm{\Psi}{\boldsymbol{\phi}}_{k}^{H}$ is the sum of independent complex normal vectors and consequently is a complex normal vector, too.
In the following, this vector is used in both LS and MMSE estimation.
\subsubsection{LS Method}
The LS estimate of ${\bm{g}}_{k}$ is derived as follows \cite{kay1993fundamentals}:
\begin{equation}\label{hhatrelationv}
\hat{\bm{g}}_{k}^{ls}=\frac{1}{{\tau}_p\sqrt{{q}_{lk}}}\bm{\Psi}{\boldsymbol{\phi}}_{k}^{H}
={\bm{g}}_{k}+\sum\limits_{j\ne l}{\sqrt{\frac{{{q}_{jk}}}{{{q}_{lk}}}}{{\bm{g}}_{ljk}}}+\frac{1}{{\tau}_p\sqrt{{q}_{lk}}}{\bm{W}}{\boldsymbol{\phi}}_{k}^{H}.
\end{equation}
In other words, the received pilot is multiplied by $\frac{1}{{\tau}_p\sqrt{{q}_{lk}}}{\boldsymbol{\phi}}_{k}^{H}$, which needs $2MK{\tau}_p$ floating point operations per coherence block.
The estimate ${\hat{\bm{g}}}_k^{ls}$ has
$\mathbb{C}\mathrm{N}\bigl( {\bm{h}}_k,{\bm{S}}_{k} \bigr)$ distribution where the mean vector~(${\bm{h}}_k$) and covariance matrix~(${\bm{S}}_{k}$) are:
\begin{align}\label{lsmean}
{{\bm{h}}_{k}}&={{\bm{m}}_{k}}+\sum\limits_{j\ne l}{\sqrt{\frac{{{q}_{jk}}}{{{q}_{lk}}}}{{\bm{m}}_{ljk}}}.\\
\label{lscov}{\bm{S}}_{k}&={{\bm{R}}_{k}}+\sum\limits_{j\ne l}{\frac{{{q}_{jk}}}{{{q}_{lk}}}{{\bm{R}}_{ljk}}}+\frac{\sigma _{n}^{2}}{{\tau}_p{q}_{lk}}\bm{I}.
\end{align}
\begin{proof}
The vector $\hat{\bm{g}}_{k}^{ls}$ is a sum of independent complex normal vectors and has a complex normal distribution.
Its mean is derived simply by taking the expectation of the right side of \eqref{hhatrelationv}.
Finally, since all the terms on the right side are cross-uncorrelated, their covariance summation equals ${\bm{S}}_{k}$.
\end{proof}
By defining LS channel estimation error as
${\tilde{\bm{g}}}_k^{ls} \triangleq {\bm{g}}_{k}-{\hat{\bm{g}}}_k^{ls}$, it has $\mathbb{C}\mathrm{N}\bigl( {\overline{\bm{h}}}_k,{\bm{T}}_k \bigr)$
distribution, where its mean vector~(${\overline{\bm{h}}}_k$) and
covariance matrix~(${\bm{T}}_k$) are as follows:
\begin{align}\label{meanlserr}
{\overline{\bm{h}}}_k&={{\bm{m}}_{k}}-{{\bm{h}}_{k}}=-\sum\limits_{j\ne l}\sqrt{\frac{{{q}_{jk}}}{{{q}_{lk}}}}{\bm{m}}_{ljk}.
\\
{\bm{T}}_k&=\sum\limits_{j\ne l}{\frac{{{q}_{jk}}}{{{q}_{lk}}}{{\bm{R}}_{ljk}}}+\frac{\sigma _{n}^{2}}{{\tau}_p{q}_{lk}}\bm{I}.\label{msels}
\end{align}
According to \eqref{meanlserr}, LS estimation is biased unless ${\bm{m}}_{ljk}=0$ for all $j \neq l$.
The estimation of the channel is considered as
the true response and
its error is incorporated into interference and noise terms.
According to \eqref{hhatrelationv},
the vectors $\hat{\bm{g}}_k^{ls}$ and $\tilde{\bm{g}}_k^{ls}$ are correlated with cross-covariance matrix
$-{\bm{T}}_k$.
The effect of this cross-correlation
on data detection will be discussed in \autoref{rate}.
\subsubsection{MMSE Method}
The MMSE estimation of $\bm{g}_k$ \cite{kay1993fundamentals} is
\begin{equation}\label{mmseformula}
\hat{\bm{g}}_{k}^{m}={\tau}_{p}\sqrt{q_{lk}}{\bm{R}}_{k}
{\boldsymbol{\Omega}}_{k}^{-1}
\left( \bm{\Psi}{\boldsymbol{\phi}}_{k}^{H}-{\tau}_{p}
\sum\limits_{j=1}^{L}\sqrt{{q}_{jk}}{\bm{m}}_{ljk} \right)
+{\bm{m}}_{k},
\end{equation}
where
\begin{equation}\label{psicov}
{\boldsymbol{\Omega}}_{k}=
{\tau}_{p}^{2}\sum\limits_{j=1}^{L}{q}_{jk}{\bm{R}}_{ljk}+
{\sigma}_{n}^{2}{\tau}_{p}\bm{I}.
\end{equation}
In contrast to computational complexity of LS estimation, MMSE method additionally has matrix-vector multiplication, as well as vector subtraction and addition, which totally needs $2M\big( MK+K{\tau}_{p}+1 \big)$ floating point operations per coherence block.
The estimation $\hat{\bm{g}}_{k}^{m}$ and its error (i.e.,
$\tilde{\bm{g}}_{k}^{m}\triangleq {{\bm{g}}_{k}}-\hat{\bm{g}}_{k}^{m}$)
have complex~normal~distribution as follows
\begin{align}\label{mmsedist}
\hat{\bm{g}}_{k}^{m}&\sim \mathbb{C}N\left( {{\bm{m}}_{k}},{\bm{U}}_k \right),
\\
\tilde{\bm{g}}_{k}^{m}&\sim \mathbb{C}N\left( \bm{0},{{\bm{V}}_{k}} \right),
\end{align}
where their covariance matrices are as follows:
\begin{align}\label{meanmmse}
{{\bm{U}}_{k}}&={\bm{R}}_{k}\bm{S}_{k}^{-1}{\bm{R}}_{k}.
\\
{{\bm{V}}_{k}}&={\bm{R}}_{k}-{\bm{U}}_{k}\label{MSEmmse}.
\end{align}
\begin{proof}
According to \eqref{mmseformula}, the vectors $\hat{\bm{g}}_{k}^{m}$ and $\bm{\Psi}{\boldsymbol{\phi}}_{k}^{H}$ have an affine relation.
Since $\bm{\Psi}{\boldsymbol{\phi}}_{k}^{H}$ is a complex normal vector, $\hat{\bm{g}}_{k}^{m}$ is a complex~normal~vector, too.
It is clear that MMSE estimation is unbiased, hence $\mathrm{E}\bigl[\hat{\bm{g}}_{k}^{m} \bigr] = \mathrm{E}\bigl[ \bm{g}_{k}\bigr]$ and $\mathrm{E}\bigl[\tilde{\bm{g}}_{k}^{m} \bigr] = 0$.
The covariance of $\hat{\bm{g}}_{k}^{m}$
(i.e., ${\bm{U}}_{k}$)
is derived by multiplying ${\tau}_{p}\sqrt{q_{lk}}{\bm{R}}_{k}
{\boldsymbol{\Omega}}_{k}^{-1}$ and its hermitian to the left and the right side of ${\boldsymbol{\Omega}}_{k}$, respectively.
This results in ${\bm{U}}_k=
{\tau}_{p}^2q_{lk}{\bm{R}}_{k}
{\boldsymbol{\Omega}}_{k}^{-1}{\bm{R}}_{k}$.
From \eqref{lscov}, it is concluded that ${\boldsymbol{\Omega}}_{k}=
{\tau}_{p}^2q_{lk}{\bm{S}}_{k}$
and the proof of \eqref{meanmmse} is now complete.
From the orthogonality of MMSE estimator and its error:
$\mathrm{E}\big[\tilde{\bm{g}}_k^m\bigl(\tilde{\bm{g}}_k^m\bigr)^H\big]=
\bigl(\mathrm{E}\big[{\bm{g}}_k{\bm{g}}_k^H\big]-
{\bm{m}}_k{\bm{m}}_k^H\bigr)-
\bigl(\mathrm{E}\big[\hat{\bm{g}}_k^m\bigl(\hat{\bm{g}}_k^m\bigr)^H\big]
-{\bm{m}}_k{\bm{m}}_k^H\bigr)$. By direct substitution, \eqref{MSEmmse} is derived.
\end{proof}
\subsection{Multi-user Data Detection}\label{dt}
We assume that each user sends data vector using the same ${\tau}_u$ time-frequency resources.
For simplicity, only one of these resources is considered.
In the data transmission phase, ${\bm{x}}_{j}\in {{\mathbb{C}}^{K}}$ is transmitted by users in the $j$th cell
such that $\mathrm{E}\bigl[ {\bm{x}}_{j}{\bm{x}}_{j}^H\bigr]=\bm{P}_{j}$,
where ${\bm{P}}_j$ is a diagonal matrix which consists of users' average powers.
The received data at the $l$th BS (i.e.,
${\bm{y}}\in {{\mathbb{C}}^{M}}$) is as follows:
\begin{equation}\label{datatransmission}
\bm{y}=\sum\limits_{i=1}^{K}{\bm{g}}_{i}{{x}_{li}}+\sum\limits_{j\ne l}{\sum\limits_{i=1}^{K}{{{\bm{g}}_{lji}}{{x}_{ji}}}}+\bm{n},
\end{equation}
where 
$x_{ji}$ is the data symbol transmitted by $i$th user in $j$th cell and
${\bm{n}}\in {{\mathbb{C}}^{M}}$ 
describes 
the noise vector consisting of i.i.d $\mathbb{C}\mathrm{N}\bigl( 0,\sigma^2_n \bigr)$ elements.
In MRC detector, the received vector ($\bm{y}$) is multiplied by $\hat{\bm{g}}_k^H$.
Thus, the detected signal of $k$th user in the $l$th cell~($\hat{x}_{k}$) is represented by:
\begin{align}
{\hat{x}}_{k}&=\hat{\bm{g}}_{k}^{H}{{\hat{\bm{g}}}_{k}}{x}_{lk}
+\sum\limits_{i=1}^{K}{\hat{\bm{g}}_{k}^{H}{{{\tilde{\bm{g}}}}_{i}}{{x}_{li}}}
+\underset{i\ne k}{\overset{{}}{\mathop \sum }}\,\hat{\bm{g}}_{k}^{H}{{\hat{\bm{g}}}_{i}}{{x}_{li}}
\nonumber\\&+\sum\limits_{j\ne l}{\sum\limits_{i=1}^{K}{\hat{\bm{g}}_{k}^{H}{{\bm{g}}_{lji}}{{x}_{ji}}}}
+\hat{\bm{g}}_{k}^{H}\bm{n}.\label{detected}
\end{align}
The detected data $\hat{x}_{k}$ includes five terms.
The first term is the desired signal.
The second one corresponds to the channel estimation error.
The third and fourth terms are intra-cell and inter-cell interference parts, respectively.
Finally, the last one is due to noise.
In the next section, this equation is used to calculate SE.
\section{Spectral efficiency in the presence of LS and MMSE channel estimation}\label{rate}
In this section, the calculation of instantaneous SE and approximating its average are discussed for imperfect CSI case.
If the desired term is uncorrelated with others,
SE of $k$th user can be obtained by ${\log}_{2}\bigl(1+{\nu}_{k}\bigr)$,
where ${\nu}_{k}$ is simply the power ratio of the desired term to undesired terms.
Since the users' data are independent, the only possibly correlated terms are
$\hat{\bm{g}}_{k}^{H}{{\hat{\bm{g}}}_{k}}{x}_{lk}$ and
$\hat{\bm{g}}_{k}^{H}{{\tilde{\bm{g}}}_{k}}{x}_{lk}$.
The vectors $\hat{\bm{g}}_{k}^{m}$ and
$\tilde{\bm{g}}_{k}^{m}$ are cross-uncorrelated while
$\hat{\bm{g}}_{k}^{ls}$ and
$\tilde{\bm{g}}_{k}^{ls}$ are not.
As a result, in the presence of LS channel estimation, signal and interference are dependent 
which makes SE analysis tricky.
To solve this, we incorporate the cross-correlated part of interference into the signal.
By extracting
$\hat{\mathrm{E}}\bigl[ \tilde{\bm{g}}_{k}\bigl| \hat{\bm{g}}_{k} \bigr. \bigr]$ (linear MMSE estimation of $\tilde{\bm{g}}_{k}$ given $\hat{\bm{g}}_{k}$)
from $\tilde{\bm{g}}_{k}$,
${\overline{\bm{g}}}_k\triangleq\tilde{\bm{g}}_{k}-\hat{\mathrm{E}}\bigl[ \tilde{\bm{g}}_{k}\bigl| \hat{\bm{g}}_{k} \bigr. \bigr]$
is uncorrelated with $\hat{\bm{g}}_{k}$.
Then, $\hat{\bm{g}}_{k}+\hat{\mathrm{E}}\bigl[ \tilde{\bm{g}}_{k}\bigl| \hat{\bm{g}}_{k} \bigr. \bigr]=\hat{\bm{g}}_{k}^{m}$,
${\overline{\bm{g}}}_k=\tilde{\bm{g}}_{k}^{m}$,
and thus \eqref{detected} can be rewritten as:
\begin{align}
{\hat{x}}_{k}
&=\hat{\bm{g}}_{k}^{H}\hat{\bm{g}}_{k}^{m}{x}_{lk}
+\sum\limits_{i\ne k}\hat{\bm{g}}_{k}^{H}\hat{\bm{g}}_{i}^{m}{{x}_{li}}
+\sum\limits_{i=1}^{K}\hat{\bm{g}}_{k}^{H}\tilde{\bm{g}}_{i}^{m}{{x}_{li}}
\nonumber\\
&+\sum\limits_{j\ne l}{\sum\limits_{i=1}^{K}{\hat{\bm{g}}_{k}^{H}{{\bm{g}}_{lji}}{{x}_{ji}}}}
+\hat{\bm{g}}_{k}^{H}{{\bm{n}}}.\label{detkthuser2}
\end{align}
The term $\hat{\bm{g}}_{k}^{H}\hat{\bm{g}}_{k}^{m}{x}_{lk}$ is uncorrelated with other terms.
Thus, the power of all expressions over a coherence block can be used for calculating
instantaneous SE.
In calculating these powers, only $\hat{\bm{g}}_{k}$ and $\hat{\bm{g}}_{k}^{m}$ are deterministic.
For any linear channel estimator, SE  of $k$th user~($\eta_{k}$) is expressed as
\begin{equation}\label{segen}
\eta_{k}=
\gamma{\log }_{2}\left( 1+\frac{{p}_{lk}
    {{\left| \hat{\bm{g}}_{k}^H\hat{\bm{g}}_{k}^{m} \right|}^{2}}}{I_1+I_2+I_3+I_4} \right),
\end{equation}
where the factor $\gamma$ equals the length ratio of uplink data to coherence block:
\begin{equation}
\gamma=\frac{{\tau}_u}{{\tau}_u+{\tau}_p}.
\end{equation}
The scalar $p_{lk}$ is power mean of $x_{lk}$.
The term ${p}_{lk}
{{\bigl| \hat{\bm{g}}_{k}^H\hat{\bm{g}}_{k}^{m} \bigr|}^{2}}$ is 
the power of $\hat{\bm{g}}_{k}^{H}\hat{\bm{g}}_{k}^{m}{x}_{lk}$.
The term $I_1$ is the power of 
$\sum_{i\ne k}\hat{\bm{g}}_{k}^{H}\hat{\bm{g}}_{i}^{m}{x}_{li} $
 and is equal to
\begin{equation}
I_1=\sum\limits_{i\ne k}{{{p}_{li}}{{\left| {{ \hat{\bm{g}}_{k} }^{H}}\hat{\bm{g}}_{i}^{m} \right|}^{2}}}.
\end{equation}
$I_2$ corresponds to the power of 
$\sum_{i=1}^{K}\hat{\bm{g}}_{k}^{H}\tilde{\bm{g}}_{i}^{m}{x}_{li}$
and is written as
\begin{equation}
I_2={{\hat{\bm{g}}_{k}}^{H}}\left( \mathop{\sum }_{i=1}^{K}{{p}_{li}}{{\bm{V}}_{i}} \right)\hat{\bm{g}}_{k}.
\end{equation}
The power of $\sum_{j\ne l}{\sum_{i=1}^{K}{\hat{\bm{g}}_{k}^{H}{{\bm{g}}_{lji}}{{x}_{ji}}}}$ is equal to
\begin{equation}
I_3=
\hat{\bm{g}}_{k}^{H}\left( \sum\limits_{j\ne l}{\sum\limits_{i=1}^{K}{{{p}_{ji}}\left( {{\bm{R}}_{lji}}+{{\bm{m}}_{lji}}\bm{m}_{lji}^{H} \right)}} \right)\hat{\bm{g}}_{k}.
\end{equation}
Finally, the term $I_4$ is the average power of $\hat{\bm{g}}_{k}^{H}{{\bm{n}}}$:
\begin{equation}
I_4=\sigma _{n}^{2}{{\left\| \hat{\bm{g}}_{k} \right\|}^{2}}.
\end{equation}
The ergodic SE is obtained by averaging \eqref{segen}
over all possible channel realizations.
Unfortunately, an exact closed-form average is not accessible.
For the ergodic SE, we propose a closed-form approximation by using the following lemma.
\begin{lem}
If $X$ and $Y$ are sums of non-negative random variables, then \cite{6816003}
\begin{equation}\label{app1+x/y}
\mathrm{E}\left[ \log \left( 1+\frac{X}{Y} \right) \right]\approx \log \left( 1+\frac{\mathrm{E}\left[ X \right]}{\mathrm{E}\left[ Y \right]} \right).
\end{equation}
It is shown in \cite{6816003} that both sides of above approximation have the same lower and also upper bounds.
By increasing the number of non-negative random variables, these bounds tighten and the approximation becomes more reliable.
\end{lem}
Hence,
\begin{equation}\label{seappgen}
\mathrm{E}\left[ \eta_{k}\right]\approx
{\overline{\eta}}_{k}=
\gamma{{\log }_{2}}\left( 1+\frac{{{p}_{lk}}\mathrm{E}\left[
    {{\left| {\hat{\bm{g}}}_{k}^{H}\hat{\bm{g}}_{k}^{m} \right|}^{2}}
     \right]}{\mathrm{E}\left[ I_1+I_2+I_3+I_4 \right]} \right).
\end{equation}
In  \cite{ozdtranc}, the expression $p_{lk}\bigl| \mathrm{E}\bigl[ {\hat{\bm{g}}}_{k}^{H}{\bm{g}}_{k}\bigr] \bigr|^2$
which equals $p_{lk}\bigl| \mathrm{E}\bigl[ {\hat{\bm{g}}}_{k}^{H}{\hat{\bm{g}}}_{k}^{m}\bigr] \bigr|^2$
is the numerator of SINR,
while in \eqref{seappgen} the numerator is
${{p}_{lk}}\mathrm{E}\bigl[
{{\bigl| {\hat{\bm{g}}}_{k}^{H}\hat{\bm{g}}_{k}^{m} \bigr|}^{2}}
\bigr]$.
As $\mathrm{E}\bigl[
{{\bigl| {\hat{\bm{g}}}_{k}^{H}\hat{\bm{g}}_{k}^{m} \bigr|}^{2}}
\bigr] \geq \bigl| \mathrm{E}\bigl[ {\hat{\bm{g}}}_{k}^{H}{\hat{\bm{g}}}_{k}^{m}\bigr] \bigr|^2$, ${\nu}_{k}$ in \eqref{seappgen} has larger numerator, as well as
smaller denominator compared to that in \cite{ozdtranc}, since they sum up to the detector output power for both works.
Therefore, it is concluded that more accurate ergodic SE can be achieved in our approach.
In the following, we compute expectations in \eqref{seappgen} for MMSE and LS estimators.
\subsection{Ergodic spectral efficiency in the presence of LS channel estimation}
In this case, the numerator in \eqref{seappgen} is calculated from the following equality:
\begin{align}\label{numofls}
\mathrm{E}\left[ {{\left| {{\left( \hat{\bm{g}}_{k}^{ls} \right)}^{H}}\hat{\bm{g}}_{k}^{m} \right|}^{2}} \right]
&=\mathrm{Tr}\left\{ \bm{R}_{k}^{2} \right\}
+\bm{m}_{k}^{H}{{\bm{S}}_{k}}{{\bm{m}}_{k}}
\nonumber\\&+\bm{h}_{k}^{H}{{\bm{U}}_{k}}\bm{h}_{k}
+{\left| \mathrm{Tr}\left\{ {{\bm{R}}_{k}} \right\}
        +{\bm{h}}_{k}^H{\bm{m}}_{k} \right|}^{2}.
\end{align}
\begin{proof}
First, $\hat{\bm{g}}_{k}^{m}$ is expressed in terms of $\hat{\bm{g}}_{k}^{l}$.
Then, \autoref{quartic} in \hyperref[quarticmom]{Appendix~2} is used.
The complete proof is provided in \hyperref[prooflsapp]{Appendix~3}.
\end{proof}
To evaluate the expectations in denominator, we define
$I_n^{ls}\triangleq \mathrm{E}\bigl[{I}_{n}\bigr]$ for $n=1,2,3,4$.
\begin{enumerate}[label=(\roman*),labelindent=\parindent,leftmargin=*]
\item
\begin{equation}\label{ea1B1}
I_1^{ls}
=\sum\limits_{i\ne k}{{p}_{li}}\mathrm{Tr}\left\{ \left( {{\bm{S}}_{k}}
+{{\bm{h}}_{k}}\bm{h}_{k}^{H} \right)
\left( {{\bm{U}}_{i}}
+{{\bm{m}}_{i}}\bm{m}_{i}^{H} \right) \right\}.
\end{equation}
\begin{proof}
The proof is given in \hyperref[proofmseapp]{Appendix~4}.
\end{proof}
\item 
\begin{equation}\label{ea2B2}
I_2^{ls}
=\sum\limits_{i=1}^{K}{{{p}_{li}}\mathrm{Tr}\left\{ {{\bm{V}}_{i}}\left( {{\bm{S}}_{k}}+{{\bm{h}}_{k}}\bm{h}_{k}^{H} \right) \right\}}.
\end{equation}
\begin{proof}
This equation is verified by calculating $\mathrm{E}\bigl[ {{\bigl( \hat{\bm{g}}_{k}^{ls} \bigr)}^{H}}{{\bm{V}}_{i}}\hat{\bm{g}}_{k}^{ls} \bigr]$ in $\mathrm{E}\bigl[ I_2\bigr]$,
which is obtained by substituting $\bm{x}=\hat{\bm{g}}_{k}^{ls}$, $\bm{m}={\bm{h}}_{k}$,
$\boldsymbol{\Sigma}={\bm{S}}_{k}$ and $\bm{A}={\bm{V}}_{i}$ in \autoref{lemquad} of \hyperref[quadmom]{Appendix~1}.
\end{proof}
\item
\begin{equation}\label{ea3B3}
I_3^{ls}
=\sum\limits_{j\ne l}\sum\limits_{i=1}^{K}{p}_{ji}\mathrm{Tr}\left\{ \left( {\bm{R}}_{lji}+{\bm{m}}_{lji}{\bm{m}}_{lji}^{H} \right)        
\left( {{\bm{S}}_{k}}+{{\bm{h}}_{k}}\bm{h}_{k}^{H} \right) \right\}.
\end{equation}
\begin{proof}
It is similar to proof of \eqref{ea2B2}, 
except for the double summation. Hence, the inner expectation is derived for specific $i$ and $j$ by replacing $\bm{A}={\bm{R}}_{lji}+{\bm{m}}_{lji}{\bm{m}}_{lji}^{H}$.
\end{proof}
\item
\begin{equation}\label{ea4B4}
I_4^{ls}
=\sigma _{n}^{2}\left( \mathrm{Tr}\left\{ {{\bm{S}}_{k}}\right\}
+\left\|{\bm{h}}_{k}\right\|^2\right).
\end{equation}
\begin{proof}
This equation is derived by substituting $\bm{x}=\hat{\bm{g}}_{k}^{ls}$, $\bm{m}={\bm{h}}_{k}$ and
$\boldsymbol{\Sigma}={\bm{S}}_{k}$ in \eqref{normmean}
of \hyperref[quadmom]{Appendix~1}.
\end{proof}
\end{enumerate}
Therefore the ergodic SE of the $k$th user~(${\overline{\eta}}_{k}^{ls}$) in the presence of LS channel estimation is approximated as
\begin{equation}\label{selsapp}
{\overline{\eta}}_{k}^{ls}=\gamma{{\log }_{2}}\left( 1+\frac{{{p}_{lk}}\mathrm{E}\left[ {{\left| {{\left( \hat{\bm{g}}_{k}^{ls} \right)}^{H}}\hat{\bm{g}}_{k}^{m} \right|}^{2}} \right]}{I_1^{ls}+I_2^{ls}+I_3^{ls}+I_4^{ls}} \right).
\end{equation}
\subsection{Ergodic spectral efficiency in the presence of MMSE channel estimation}
In this case, either of \eqref{detected} or \eqref{detkthuser2} can be used for calculating $ {\nu}_{k} $, because $\hat{\mathrm{E}}\bigl[ \tilde{\bm{g}}_{k}^{m}\bigl| \hat{\bm{g}}_{k}^{m} \bigr. \bigr]=\bm{0}$.
If ${\hat{\bm{g}}}_{k}$ is replaced by $\hat{\bm{g}}_{k}^{m}$ in \eqref{seappgen},
then the numerator becomes
\begin{equation}\label{numofmmse}
    \mathrm{E}{\left[ {\left\| \hat{\bm{g}}_{k}^{m} \right\|}^{4}\right] }=\mathrm{Tr}\left\{ \bm{U}_{k}^{2} \right\}+2\bm{m}_{k}^{H}{{\bm{U}}_{k}}{{\bm{m}}_{k}}+{{\left( \mathrm{Tr}\left\{ {{\bm{U}}_{k}} \right\}+{\left\|{\bm{m}}_{k}\right\|}^2 \right)}^{2}}.
\end{equation}
\begin{proof}
    See \hyperref[quarticmom]{Appendix~2}.
\end{proof}
If we denote the means of $I_1$, $I_2$, $I_3$ and $I_4$ over all possible realizations of the channel by
$I_1^{m}$, $I_2^{m}$, $I_3^{m}$ and $I_4^{m}$ respectively:
\begin{enumerate}[label=(\roman*),labelindent=\parindent,leftmargin=*]
    \item 
    \begin{equation}\label{ea1A1}
        I_1^{m}=\mathop{\sum }_{i\ne k}{{p}_{li}}\mathrm{Tr}\left\{ \left( {{\bm{U}}_{k}}+{{\bm{m}}_{k}}\bm{m}_{k}^{H} \right)\left( {{\bm{U}}_{i}}+{{\bm{m}}_{i}}\bm{m}_{i}^{H} \right) \right\}.
    \end{equation}
    \begin{proof}
The proof has the same procedure for \eqref{ea1B1},
except that $\hat{\bm{g}}_{k}^{ls}$ is replaced with $\hat{\bm{g}}_{k}^{m}$.
Consequently, ${\bm{h}}_{k}$ and ${\bm{S}}_{k}$ are replaced with ${\bm{m}}_{k}$ and ${\bm{U}}_{k}$, respectively.
    \end{proof}
    \item
    \begin{equation}\label{ea2A2}
        I_2^{m}=
        \sum\limits_{i=1}^{K}{p}_{li}\mathrm{Tr}\left\{ {\bm{V}}_{i}\left( {\bm{U}}_{k}+{\bm{m}}_{lk}\bm{m}_{k}^{H} \right) \right\}.
    \end{equation}
    \begin{proof}
        It is identical to \eqref{ea2B2} proof except for
        $\bm{x}=\hat{\bm{g}}_{k}^{m}$, $\bm{m}={\bm{m}}_{k}$,
        $\boldsymbol{\Sigma}={\bm{U}}_{k}$ and $\bm{A}={\bm{V}}_{i}$.
    \end{proof}
    \item
    \begin{equation}\label{ea3A3}
        I_3^{m}=\sum\limits_{j\ne l}\sum\limits_{i=1}^{K}{{p}_{ji}}
        \mathrm{Tr}\left\{ 
        \left({\bm{R}}_{lji}+{\bm{m}}_{lji}{\bm{m}}_{lji}^{H}\right)
        \left( {{\bm{U}}_{k}}+{{\bm{m}}_{k}}\bm{m}_{k}^{H} \right)    
        \right\}.
    \end{equation}
    \begin{proof}
        It is identical to the proof of \eqref{ea3B3}. The only difference is that
        ${\bm{h}}_{k}$ and ${\bm{S}}_{k}$ 
        are replaced with ${\bm{m}}_{k}$ and ${\bm{U}}_{k}$, respectively.
    \end{proof}
    \item
    \begin{equation}\label{ea4A4}
        I_4^{m}
        =\sigma _{n}^{2}\left( \mathrm{Tr}\left\{ {{\bm{U}}_{k}}\right\}+\left\|{\bm{m}}_{k}\right\|^2\right).
    \end{equation}
    \begin{proof}
        From \eqref{normmean} in \hyperref[quadmom]{Appendix~1}, we have
        $\mathrm{E}\bigl[ {{\bigl\| \hat{\bm{g}}_{k}^{m} \bigr\|}^{2}} \bigr]= \mathrm{Tr}\bigl\{ {{\bm{U}}_{k}} \bigr\}+{{\bigl\| {{\bm{m}}_{k}} \bigr\|}^{2}}$, which concludes \eqref{ea4A4}.
    \end{proof}
\end{enumerate}
Therefore, the approximated ergodic SE of the $k$th user~(${\overline{\eta}}_{k}^{mmse}$) in the presence of MMSE channel estimation is expressed as
\begin{equation}\label{semmseapp}
    {\overline{\eta}}_{k}^{mmse}=
    \gamma{{\log }_{2}}\left( 1+\frac{{{p}_{lk}}\mathrm{E}{\left[ {\left\| \hat{\bm{g}}_{k}^{m} \right\|}^{4}\right] }}{I_1^{m}+I_2^{m}+I_3^{m}+I_4^{m}} \right).
\end{equation}
\subsection{Some Remarks on Impact of Spatial Correlation}\label{impacscor}
In this subsection, the impact of spatial correlation is investigated on SE performance. 
The common belief is that the spatial correlation degrades SE, which is true for single-user MIMO communications.
However, it is shown for Rayleigh channel that sometimes it does not happen in multi-user MIMO \cite{massivemimobook,5353268}.
We show that it is theoretically possible for multi-user MIMO correlated Rician channel to have a greater SE than uncorrelated in some conditions.

For this purpose, the approximation in \eqref{selsapp} or \eqref{semmseapp} is compared in terms of the covariance matrix for both channels.
The covariance matrix of the uncorrelated channel (${\bm{R}}_{ljk}^{uc}$) is derived by zeroing off-diagonal elements in that of the correlated one (${\bm{R}}_{ljk}$).
Since the logarithm is an increasing function, we compare the inside argument in \eqref{selsapp} or \eqref{semmseapp} instead of SE.
Equations \eqref{selsapp} and \eqref{semmseapp} depend on covariance matrix of MMSE estimation (${\bm{U}}_{k}$), which needs matrix inversion according to \eqref{meanmmse}.
Therefore, the comparison is complicated.

To simplify the problem, the perfect CSI condition is considered in which no channel estimation error exists.
In this case, ${\hat{\bm{g}}}_k^{m}={\hat{\bm{g}}}_k^{ls}=\bm{g}_k$.
Therefore the same result is obtained by using either of \eqref{selsapp} or \eqref{semmseapp} equations.
The arguments inside the logarithms also become equal to that of the perfect CSI case.
We subject our analysis to \eqref{semmseapp}, which corresponds for SE of MMSE channel estimator.
Therefore, in this ideal condition we have $\bm{U}_k=\bm{R}_k$ and $\bm{V}_k=0$.
By substituting these matrices in \eqref{numofmmse} to
\eqref{ea4A4},
the term \eqref{ea2A2} becomes equal to zero for both correlated and uncorrelated channels since no estimation error is assumed.
Here, the equation \eqref{ea4A4} has the same value for both channels, because of $\mathrm{Tr}\bigl\{ {{\bm{R}}_{k}} \bigr\}=\mathrm{Tr}\bigl\{ {\bm{R}}_{k}^{uc} \bigr\}$.

The expressions inside \eqref{semmseapp} are compared peer to peer for both correlated and uncorrelated Rician channels.
In the nominator of the fraction, the terms $\mathrm{Tr}\big\{ \bm{R}_{k}^{2} \big\}$ and, $\bm{m}_{k}^{H}{{\bm{R}}_{k}}{{\bm{m}}_{k}}$ and in the denominator the term
$\big( {{\bm{R}}_{k}}+{{\bm{m}}_{k}}\bm{m}_{k}^{H} \big)\big( {{\bm{R}}_{ljk}}+{{\bm{m}}_{ljk}}\bm{m}_{ljk}^{H} \big)$ have different values for correlated and uncorrelated channels. We compare them as follows:
\begin{rem}
    \label{trRk2}
According to \eqref{trA2} in \hyperref[offdiag]{Appendix~5}, it is concluded that
$\mathrm{Tr}\big\{ \bm{R}_{k}^{2} \big\}>\mathrm{Tr}\big\{ {\big(\bm{R}_{k}^{uc}\big)}^2 \big\} $ for all cases.
\end{rem}
\begin{rem}\label{mkHRkmk}
According to \eqref{xHAx} in \hyperref[offdiag]{Appendix~5}, comparison of 
$\bm{m}_{k}^{H}{{\bm{R}}_{k}}{{\bm{m}}_{k}}$ and
$\bm{m}_{k}^{H}{{\bm{R}}_{k}^{uc}}{{\bm{m}}_{k}}$ depends on the sign of term containing real part.
If a less practical scenario is assumed such that LOS and non-LOS paths between a user and BS have the same angle of arrival, this term becomes positive. Consequently, $\bm{m}_{k}^{H}{{\bm{R}}_{k}}{{\bm{m}}_{k}}>\bm{m}_{k}^{H}{{\bm{R}}_{k}^{uc}}{{\bm{m}}_{k}}$.
Furthermore, by considering $\mathrm{Tr}\big\{ \bm{R}_{k}^{2} \big\}>\mathrm{Tr}\big\{ {\big(\bm{R}_{k}^{uc}\big)}^2 \big\} $, signal power is greater for the correlated channels in this improbable scenario.
\end{rem}
\begin{rem}\label{mkHRkmk0}
Note that in Rayleigh channel where ${\bm{m}}_{k}=0$, we have
$\bm{m}_{k}^{H}{{\bm{R}}_{k}}{{\bm{m}}_{k}}=\bm{m}_{k}^{H}{{\bm{R}}_{k}^{uc}}{{\bm{m}}_{k}}=0$. Thus correlated Rayleigh channel has higher signal power than the uncorrelated one.
\end{rem}
\begin{rem}
    \label{mkHRkmkn}
In the Rician correlated channel, if the phase difference between corresponding elements of ${\bm{R}}_k$ and ${\bm{m}}_{k}{\bm{m}}_{k}^{H}$ is between $\frac{\pi}{2}$ and $\frac{3\pi}{2}$, then $\bm{m}_{k}^{H}{{\bm{R}}_{k}}{{\bm{m}}_{k}}<\bm{m}_{k}^{H}{{\bm{R}}_{k}^{uc}}{{\bm{m}}_{k}}$ and vice versa.
\end{rem}
\begin{rem}\label{intcomp} The interference power comparison of the correlated and uncorrelated channels depends on the sign of the real part in equations \eqref{xHAx} and \eqref{trAB} in \hyperref[offdiag]{Appendix~5}.
\end{rem}
\begin{rem}
\label{idencov}If the matrices (${{\bm{R}}_{ljk}}+{{\bm{m}}_{ljk}}\bm{m}_{ljk}^{H}$) are scaled version of each other, the correlated channel case has higher interference power than the uncorrelated one.
\end{rem}
\begin{rem}\label{orthcov}
If we consider an extreme case that matrices (${{\bm{R}}_{ljk}}+{{\bm{m}}_{ljk}}\bm{m}_{ljk}^{H}$) are orthogonal to each other, the correlated channel case has zero interference power.
\end{rem}
If we consider the Rayleigh channel and assume that the matrices ${{\bm{R}}_{ljk}}$ are orthogonal, based on the \autoref{mkHRkmk0} and \autoref{orthcov} the correlated channel has higher signal and lower interference power. Hence it has a greater SE.

In the correlated Rician channel, if the matrices (${{\bm{R}}_{ljk}}+{{\bm{m}}_{ljk}}\bm{m}_{ljk}^{H}$) are orthogonal and 
LOS and non-LOS paths of each user have the same angles of arrivals,
based on the \autoref{mkHRkmk} and \autoref{orthcov}, this channel has lower interference power, as well as higher signal power.
Consequently, in this case, the SE of the correlated Rician channel is higher than the uncorrelated one.
As a result, theoretically, SE of a correlated Rician channel can be higher than the uncorrelated one in some extreme cases.
\section{Numerical Results}\label{sim}
In this section, approximations in \eqref{semmseapp} and \eqref{selsapp} are compared with 
experimental average of \eqref{segen} derived by Monte Carlo simulation results.
They are also compared with lower bounds in \cite{ozdtranc}.
The SE curves related to their work are obtained by running their MATLAB functions.
In all multi-cell simulations, a system with $L=16$ cells
is considered.
Two different cell structures have been examined as shown in \autoref{fig:comp} (a square gird) and \autoref{fig:16cellrandgeom} (a stochastic grid).
The results in \autoref{fig:comp} to \autoref{fig:mcell_uray_20190629} are related to the square grid structure and \autoref{fig:mcellcricerandbs20191205} and \autoref{fig:sevsk} are based on stochastic grid.
It is assumed that $K=10$ users are serviced in each cell.
\begin{figure}
    \centering
    \includegraphics[width=.75\linewidth]{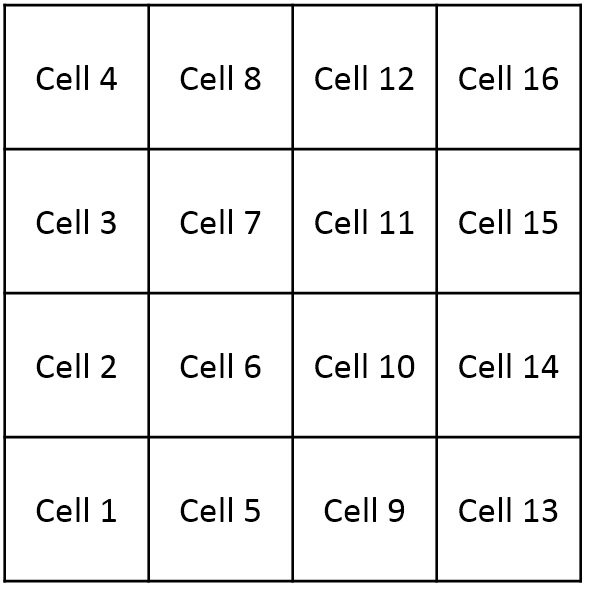}
    \caption{Configuration of $L=16$ multi-cell system with square grid.}
    \label{fig:16cell}
\end{figure}
\subsection{Parameters}
As mentioned in \autoref{model}, ${\bm{R}}_{ljk}$ depends on 
non-LOS large scale fading multiple (${\beta}_{ljk}^{NLOS}$)
and ${\bm{m}}_{ljk}$ on LOS large scale fading multiple (${\beta}_{ljk}^{LOS}$).
If a strong LOS component exists then
\begin{align}
    {\beta}_{ljk}^{LOS}&=\frac{{\kappa}_{ljk}}{1+{\kappa}_{ljk}}{\beta}_{ljk},\\
    {\beta}_{ljk}^{NLOS}&=\frac{1}{1+{\kappa}_{ljk}}{\beta}_{ljk},
\end{align}
where ${\beta}_{ljk}$ (total large scale fading multiple) and
${\kappa}_{ljk}$ (Rician factor) are defined as follows \cite{ETSIMIMO}: 
\begin{equation}
{\beta}_{ljk}\left[ \text{dB}\right] =-30.18-26{\log}_{10}\left( r_{ljk}\right)+z_{ljk},\, z_{ljk}\sim \mathrm{N}\left( 0,16\right),
\end{equation}
\begin{equation}
    {\kappa}_{ljk}\left[\text{dB}\right]=13-0.03{r}_{ljk}, 
\end{equation}
where $r_{ljk}$ is the distance (in meter) between the $k$th user in $j$th cell and BS in the $l$th cell and $z_{ljk}$ is a related shadowing factor.

Some channels may have only non-LOS components, especially when there is a large distance between user and BS.
If no LOS exists, then ${\beta}_{ljk}^{LOS}=0$ and $ {\beta}_{ljk}^{NLOS}={\beta}_{ljk}$.
In this case, we have:
\begin{equation}
{\beta}_{ljk}\left[ \text{dB}\right] =-34.53-38{\log}_{10}\left( r_{ljk}\right)+z_{ljk},\, z_{ljk}\sim \mathrm{N}\left( 0,100\right).
\end{equation}
LOS path existence depends on $r_{ljk}$.
In the assumed model, no LOS exists for $r_{ljk}\geq 300$ m, and in other cases, it exists with the probability of $1-r_{ljk}/300$.
The $s$th element of ${\bm{m}}_{ljk}$ is modelled as \cite{massivemimobook}
\begin{equation}
\left[ {\bm{m}}_{ljk}\right]_s=
\sqrt{{\beta}_{ljk}^{LOS}}
\exp\left(J\pi \left( s-1\right) \sin({\theta}_{ljk})\right), 
\end{equation}
where $J\triangleq \sqrt{-1}$
and ${\theta}_{ljk}$ denotes the angle of arrival at $l$th BS from $k$th user in $j$th cell.
This parameter is assumed to be uniformly distributed between $ 0 $ and $2\pi$.
Note in this model, the antenna spacing is a half wavelength.
The $\bigl( s,t\bigr)$th element of ${\bm{R}}_{ljk}$ is presented as \cite{massivemimobook}:
\begin{align}
\left[ {\bm{R}}_{ljk}\right]_{s,t}&=
\frac{{\beta}_{ljk}^{NLOS}}{N}\sum_{n=1}^{N}\exp\left( J\pi\left( s-t\right)\sin \left({\theta}_{ljk}^{n}\right)\right)\nonumber\\
&\times\exp\left( -\frac{\sigma_{\theta}^2}{2}\left( \pi\left( s-t\right)\cos\left({\theta}_{ljk}^{n}\right)\right)^2\right),\label{CorrCov}
\end{align}
where $\sigma_{\theta}^2$ is the angular variance and
${\theta}_{ljk}^{n}$ is angle of arrival from the $n$th scattering cluster.
For the uncorrelated channel, it is assumed that
\begin{equation}\label{UncorrCov}
{\bm{R}}_{ljk}^{u}={\beta}_{ljk}^{NLOS}\bm{I}.
\end{equation}
In our calculations, the data and pilot symbol powers ($p_{lk}$ and $q_{lk}$) are different for all $l$ and $k$. However, we assume $p_{lk}=q_{lk}$ in all simulations.
These values are chosen such that $p_{lk}{\beta}_{llk}$ becomes fixed for all $k$.
Moreover, the maximum value of $p_{lk}$ is 10~dBm in each cell.
Finally, the orthogonal pilot sequences are generated by discrete Fourier transform~(DFT) basis i.e.
\begin{equation}
    \left[ {\boldsymbol{\phi}}_{k}\right]_{s}=\exp\left( \frac{J2\pi\left(k-1\right)\left( s-1\right)}{{\tau}_{p}}\right).
\end{equation}
The values of some important parameters are listed in \autoref{param}.
\begin{table}[!t] \caption{Simulation Parameters.}\label{param}
    \centering
    \begin{tabular}{|>{\em}m{.45\columnwidth}|c|}
        \hline
        \normalfont{\textbf{Parameter}} & \textbf{Value}\\ \hline\hline
        Number of Channel Realizations& 100\\ \hline
        Number of Symbols in a Coherence Block (${\tau}_u+{\tau}_p$)& 200\\
        \hline
        Number of Each User's Pilot Symbols ($\tau_p$)&10 \\ \hline
        Noise Power ($\sigma^2_n$)& -94 dBm\\ \hline
        Maximum User Transmitted Power & 10~dBm \\
        \hline
        Angle of Arrival (${\theta}_{ljk}$) & $ {\theta}_{ljk}\sim\mathbb{U}\bigl( 0,2\pi\bigr) $\\
        \hline
        Angular Standard Deviation ($\sigma_{\theta}^2$) & $10^{\circ}$ \\ \hline
Number of Scattering Clusters ($N$) & 10 \\ \hline
Angle of Arrival from the $n$th Cluster (${\theta}_{ljk}^{n}$)  & $ {\theta}_{ljk}^{n}\sim\mathbb{U}\bigl( {\theta}_{ljk}-40^{\circ}, {\theta}_{ljk}+40^{\circ}\bigr) $\\ \hline
    \end{tabular}
\end{table}
\subsection{Results}
\begin{figure}[t!]
    \begin{subfigure}{\linewidth}
    \centering
    \includegraphics[width=\linewidth]{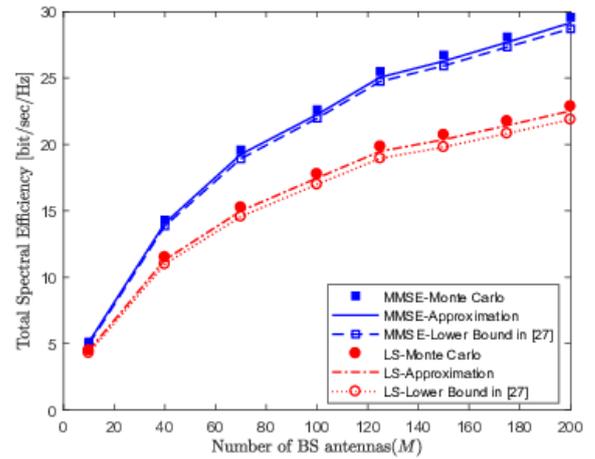}
    \caption{Multi-cell Rician Channel.}
    \label{fig:mcell_crice_20190629}
    \end{subfigure}   
\begin{subfigure}{\linewidth}
    \centering
    \includegraphics[width=\linewidth]{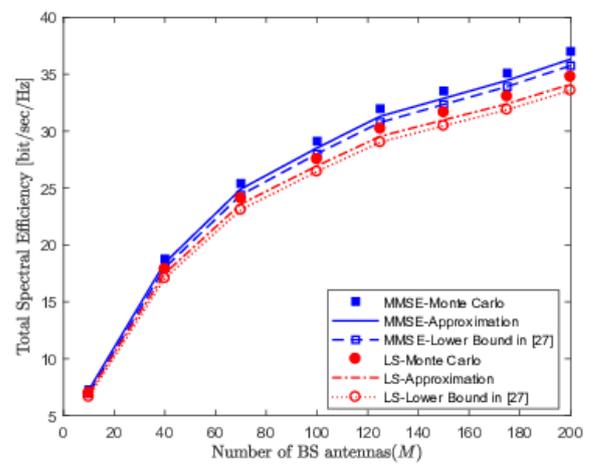}
    \caption{Single-cell Rician Channel.}
    \label{fig:scell_crice_20190629}
\end{subfigure}
\begin{subfigure}{\linewidth}
\centering
\includegraphics[width=\linewidth]{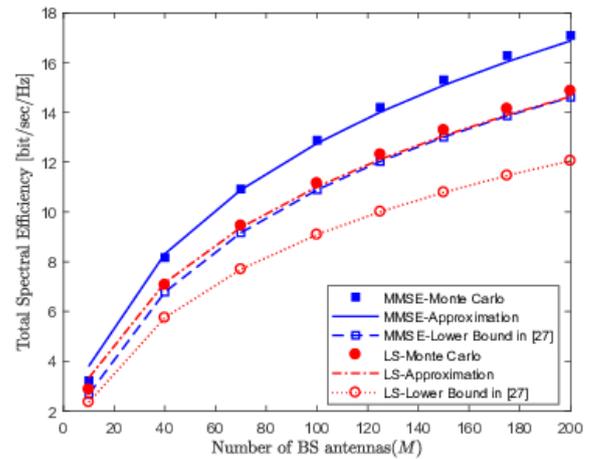}
\caption{Multi-cell Rayleigh Channel.}
\label{fig:mcell_cray_20190629}
\end{subfigure}
    \caption{Comparison of our SE approximations with corresponding Monte Carlo simulations and lower bounds in a system with a correlated channel for MRC detector and MMSE and LS channel estimators.}
    \label{fig:comp}
\end{figure}

It is assumed in square grid structure that each BS is placed in the middle of the corresponding cell.
Also in this grid, users are located randomly within a radius of 35 m to 250 m from BS in each cell.
In \autoref{fig:mcell_crice_20190629}, our proposed SE is compared with Monte Carlo simulation results
and also with lower bounds in \cite{ozdtranc} for a multi-cell massive MIMO system with correlated Rician channel.
Our approximation and simulation results are close to each other for both LS and MMSE channel estimators.
It is also seen that our SE is above the lower bound in \cite{ozdtranc}.

SE of a single cell system is provided in \autoref{fig:scell_crice_20190629}.
No pilot contamination and inter-cell interference exist in this system.
It is seen that more SE can be achieved in a single cell system compared to the multi-cell system because only intra-cell interference exists.
The difference between corresponding curves in \autoref{fig:mcell_crice_20190629} and \autoref{fig:scell_crice_20190629} is higher for the LS channel estimation case.
It can be concluded that the LS estimator is more sensitive to pilot contamination than MMSE.

In \autoref{fig:mcell_cray_20190629} performance result for correlated Rayleigh channel is considered,
where ${\bm{m}}_{ljk}$ is zero.
In this case, non-LOS large scale fading is the same as that in the channel model of \autoref{fig:mcell_crice_20190629},
which causes a decrease in both signal and interference powers.
In \autoref{fig:mcell_cray_20190629}, less SE is seen than in \autoref{fig:mcell_crice_20190629}.
It can be concluded that in this case, the signal power has been decreased more than the interference.

\begin{figure}[!t]
    \begin{subfigure}{\linewidth}
    \centering
    \includegraphics[width=\linewidth]{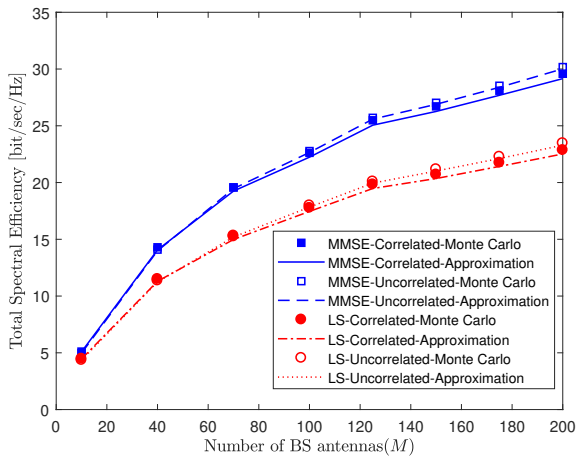}
    \caption{Rician Channel.}
    \label{fig:mcell_rice_201912}
    \end{subfigure}
\begin{subfigure}{\linewidth}
    \centering
    \includegraphics[width=\linewidth]{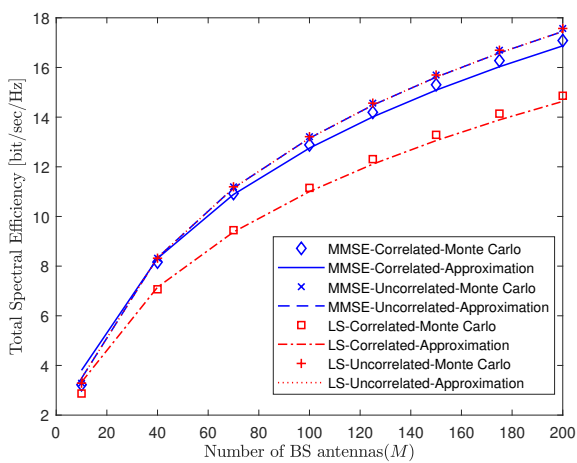}
    \caption{Rayleigh Channel.}
    \label{fig:mcell_ray_20190724}
\end{subfigure}
\begin{subfigure}{\linewidth}
\centering
\includegraphics[width=\linewidth]{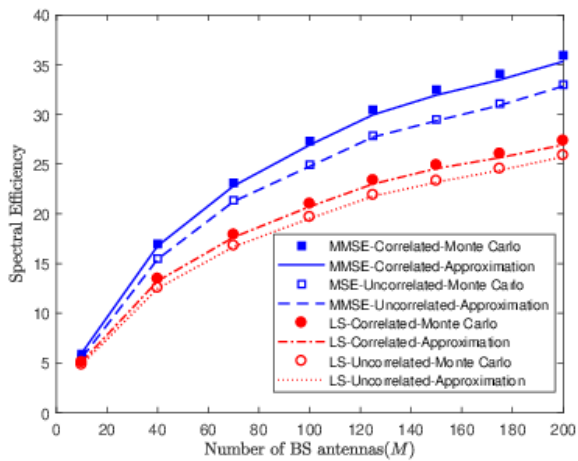}
\caption{Rician Channel by ${\theta}_{ljk}^{n}={\theta}_{ljk}$ assumption.}
\label{fig:mcell_rice_20190724}
\end{subfigure}
\caption{SE Comparison of correlated and uncorrelated channel in a multi-cell system for MRC detection and LS and MMSE channel estimation.}
\label{fig:corrVSuncor}
\end{figure}

\autoref{fig:mcell_rice_201912} shows SE performance comparison for both correlated and uncorrelated Rician channel in the presence of LS and MMSE estimations.
In this case, the uncorrelated channel curves are above the correlated ones at the high number of BS antennas.
However, at low values of $M$, the correlated channel shows slightly higher SE compared to the uncorrelated one.
In other simulations for a highly correlated channel, we observed that SE is much lower than the corresponding curves in this Figure.

\begin{figure}[!t]
    \centering
    \includegraphics[width=\linewidth]{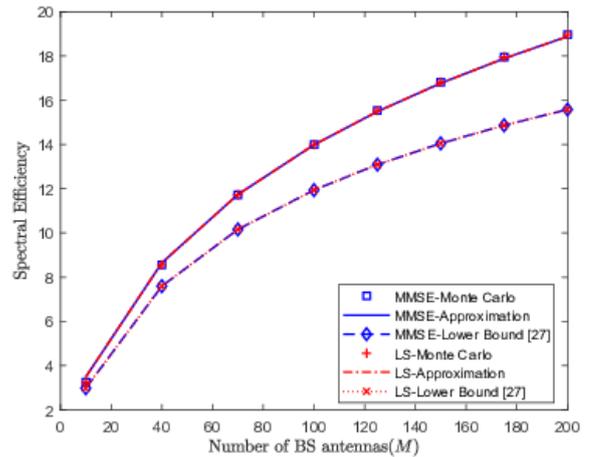}
    \caption{Comparison of our SE approximations with corresponding Monte Carlo simulations and lower bounds in a multi-cell system with uncorrelated Rayleigh channel for MRC detector and MMSE and LS channel estimators.}
    \label{fig:mcell_uray_20190629}
\end{figure}

In \autoref{fig:mcell_ray_20190724}, the same comparison is done for the Rayleigh channel.
Here, higher SE is seen for MMSE estimation in the correlated case.
For LS estimation, SE of the uncorrelated channel is above the correlated channel.
However, at the low number of antennas, it is vice versa for MMSE estimation.
Moreover, it is seen that LS and MMSE estimators have the same SE in uncorrelated Rayleigh channel, as well as the lower bounds \cite{ozdtranc} (as in \autoref{fig:mcell_uray_20190629}).
That is because here ${\hat{\bm{g}}}_{k}^{m}$ is a scale of ${\hat{\bm{g}}}_{k}^{ls}$
and this scale does not affect $ {\nu}_{k} $ in both works.
In \autoref{fig:mcell_cray_20190629} and \autoref{fig:mcell_ray_20190724} an obvious gap between Monte~Carlo and approximation curves is seen at $M=10$.
That is because the approximation in \eqref{app1+x/y} is less accurate at the low number of BS antennas.

In \autoref{fig:mcell_rice_20190724}, the assumption ${\theta}_{ljk}^{n}={\theta}_{ljk}$ is considered for Rician channel, which is improbable in practice.
As it is seen, the curves related to correlated case are above the corresponding uncorrelated curves.
As discussed in \autoref{impacscor}, the correlated channel has greater signal power than uncorrelated in this scenario.
Depending on the matrices ${{\bm{R}}_{ljk}}+{{\bm{m}}_{ljk}}\bm{m}_{ljk}^{H}$, the interference of the correlated channel can also be lower than the uncorrelated one.
Therefore,  in some cases, it is reasonable to have better performance for the correlated channel.

\begin{figure}
    \centering
    \includegraphics[width=\linewidth]{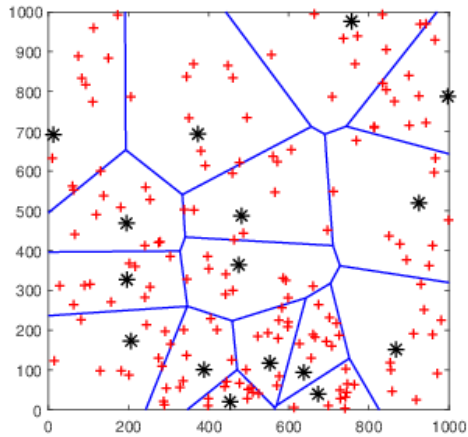}
    \caption{A stochastic grid with $L=16$ cell and randomly located BS's and users. BS: * points; Users: + points; Cell boundary: blue line.}
    \label{fig:16cellrandgeom}
\end{figure}

\begin{figure}
    \centering
    \includegraphics[width=\linewidth]{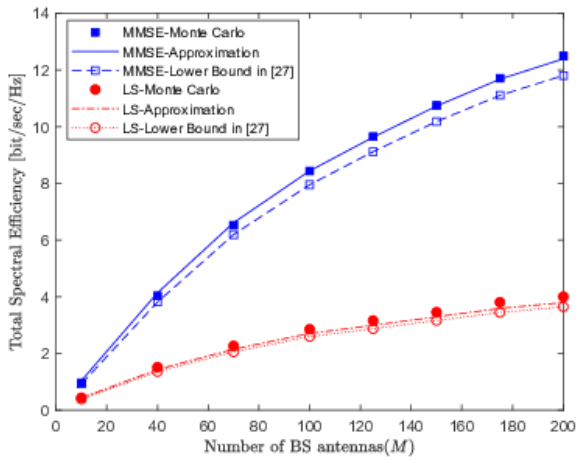}
    \caption{Total SE of the system in \autoref{fig:16cellrandgeom} by assuming correlated Rician channel.}
    \label{fig:mcellcricerandbs20191205}
\end{figure}

In the next simulation, a system is considered with a $L=16$ Voronoi cell structure \cite{haenggi_2012}.
It is assumed that all BS's are randomly located in a square area with 1000 m length, and users are serviced by the nearest BS in each cell.
An example of such a system is shown in \autoref{fig:16cellrandgeom}, where $K=10$ users are serviced in each cell.
The other settings of this simulation are the same as previous simulations.
By considering this stochastic geometry, the total SE is derived as in \autoref{fig:mcellcricerandbs20191205}.
Our approximation is still above the lower bounds in \cite{ozdtranc}.
At the low number of BS antennas, the gap between these curves is tiny but by increasing the number of the antennas, it increases.

\begin{figure}
    \centering
    \includegraphics[width=\linewidth]{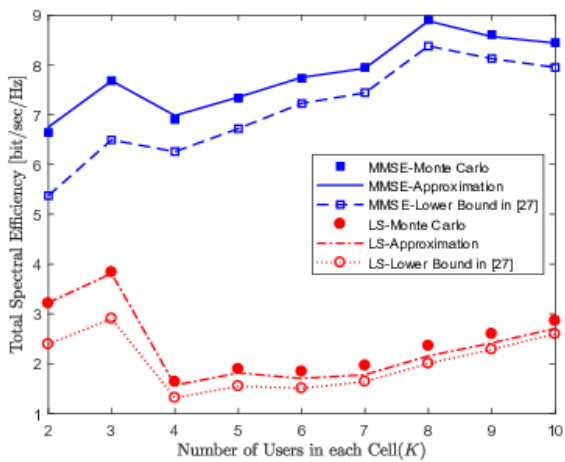}
    \caption{Total SE of MRC detection for $M=100$ antennas versus the number of users in the presence of LS and MMSE channel estimation and correlated Rician channel.}
    \label{fig:sevsk}
\end{figure}

In \autoref{fig:sevsk}, total SE is sketched in terms of $K$ for $M=100$ in a correlated Rician channel.
Since adding new users increases the power of interference, SE of other users is decreased.
On the other hand, SE of the new user is added to the total SE.
Consequently, no monotonic curve is seen.
In \autoref{fig:sevsk}, our approximations are close to the corresponding Monte Carlo simulation results.
It is also observed that the approximations are still higher than the lower bounds in \cite{ozdtranc}. However, a small gap is seen between the corresponding curves.
In most of the simulations, it is seen that the approximations in \eqref{selsapp} and \eqref{semmseapp} are close enough to Monte Carlo simulation results of \eqref{segen}.
In all simulations, it is seen that the proposed approximations are above the lower bounds provided in \cite{ozdtranc}.
\section{Conclusion}\label{conc}
In this work, instantaneous SE was analysed for multi-cell massive MIMO with correlated Rician channel.
Different from previous works, instantaneous SE is calculated based on a new  SINR derivation.
Then, closed~form approximations were proposed for ergodic SE of MRC detector in the presence of LS and MMSE estimators.
These approximations can be used for SE optimisation, as well as energy efficiency analysis.
It was shown that our approximations are close to the Monte Carlo simulation results at the high number of antennas and also higher than the lower bounds in earlier works.
Furthermore, it was observed that MMSE channel estimation outperforms LS, or they have the same SE in uncorrelated Rayleigh case.
In future work, we consider applying our approach to ZF or MMSE detector to obtain the instantaneous SE. Moreover, we investigate the ergodic SE approximation of these detectors.

\bibliographystyle{iet}
\bibliography{Revision2-references}
\section{Appendices}
\subsection{Appendix 1: Quadratic Expectations of a Complex Normal Vector}\label{quadmom}\def\subsectionautorefname{Appendix}
\begin{lem}\label{lemquad}
If $\bm{x}\sim \mathbb{C}N\bigl( \bm{m},\boldsymbol{\Sigma} \bigr)$ then for any
matrix $\bm{A}$ 
we have
\begin{equation}\label{qudratic}
\mathrm{E}\left[\bm{x}^H\bm{A}\bm{x}\right]=\mathrm{Tr}\left\{ \bm{A}\bigl(\boldsymbol{\Sigma}+ \bm{m}{\bm{m}}^H \bigr) \right\}.
\end{equation}
\end{lem}
In an especial case, that $\bm{A}=\bm{I}$, \eqref{qudratic} is simplified as
\begin{equation}\label{normmean}
\mathrm{E}\bigl[{\left\| \bm{x}\right\|}^2 \bigr]=\mathrm{Tr}\left\lbrace \boldsymbol{\Sigma}\right\rbrace + {\left\| \bm{m}\right\| }^2.
\end{equation}
Another quadratic expectation used in our derivations is
\begin{equation}\label{quadmat}
\mathrm{E}\left[ \bm{x}{\bm{x}}^H\right] = \boldsymbol{\Sigma}+\bm{m}{\bm{m}}^H.
\end{equation}
\subsection{Appendix 2: Quartic Expectation of a Complex Normal Vector}\label{quarticmom}
\begin{lem}\label{quartic}
    If $\bm{x}\sim \mathbb{C}N\bigl( \bm{m},\boldsymbol{\Sigma} \bigr)$ then for any non-zeros
    matrices $\bm{A}$, $\bm{B}$, $\bm{C}$, $\bm{D}$ and 
    arbitrary vectors 
    $\bm{a}$, $\bm{b}$, $\bm{c}$, $\bm{d}$
    the following expectation holds \cite{matrixref}
    \begin{align}\label{qurt}
    & \mathrm{E}\left[ {{\left( \bm{Ax}+\bm{a} \right)}^{H}}\left( \bm{Bx}+\bm{b} \right){{\left( \bm{Cx}+\bm{c} \right)}^{H}}\left( \bm{Dx}+\bm{d} \right) \right]=\nonumber\\&\mathrm{Tr}\left\{ {\bm{A}}^{H}\bm{B}\boldsymbol{\Sigma} {{\bm{C}}^{H}}\bm{D}\boldsymbol{\Sigma}\right\} 
    +{{\left( \bm{Cm}+\bm{c} \right)}^{H}}\bm{D}\boldsymbol{\Sigma}{{\bm{A}}^{H}}\left( \bm{Bm}+\bm{b} \right) \nonumber\\ 
    & +{{\left( \bm{Am}+\bm{a} \right)}^{H}}\bm{B}\boldsymbol{\Sigma}{{\bm{C}}^{H}}\left( \bm{Dm}+\bm{d} \right) \nonumber\\ 
    & +\left( \mathrm{Tr}\left\{ \bm{B}\boldsymbol{\Sigma}{{\bm{A}}^{H}} \right\}+{{\left( \bm{Am}+\bm{a} \right)}^{H}}(\bm{Bm}+\bm{b}) \right) \nonumber\\ 
    & \times \left( \mathrm{Tr}\left\{ \bm{D}\boldsymbol{\Sigma}{{\bm{C}}^{H}} \right\}+{{\left( \bm{Cm}+\bm{c} \right)}^{H}}\left( \bm{Dm}+\bm{d} \right) \right).
    \end{align}
    For a specific case that $\bm{A}=\bm{B}=\bm{C}=\bm{D}=\bm{I}$ and
    $\bm{a}=\bm{b}=\bm{c}=\bm{d}=\bm{0}$, it follows that
    \begin{equation}\label{4mom}
\mathrm{E}\left[\left\| \bm{x}\right\|^4 \right] =
\mathrm{Tr}\left\lbrace \boldsymbol{\Sigma}^{2} \right\rbrace+2{\bm{m}}^{H}\boldsymbol{\Sigma}\bm{m}+{\left( \mathrm{Tr}\left\{ \boldsymbol{\Sigma} \right\}+{\left\| \bm{m} \right\|}^{2} \right)}^{2}.
    \end{equation}
\end{lem}
\subsection{Appendix 3: Equation \eqref{numofls} Proof}\label{prooflsapp}
For calculating $\mathrm{E}\bigl[ {{\bigl| \hat{\bm{g}}_{k}^{ls}\hat{\bm{g}}_{k}^{m} \bigr|}^{2}} \bigr]$,
MMSE estimation~($\hat{\bm{g}}_{k}^{m}$) is rewritten in terms of $\hat{\bm{g}}_{k}^{ls}$ as follows:
\begin{equation}\label{mmsels}
\hat{\bm{g}}_{k}^{m}={\bm{F}}_{k}\hat{\bm{g}}_{k}^{ls}+{{\bm{f}}_{k}},
\end{equation}
where
\begin{align}\label{dkdef}
{\bm{F}}_{k}&={{\bm{R}}_{k}}\bm{S}_{k}^{-1},
\\
{\bm{f}}_{k}&={\bm{m}}_{k}-{\bm{F}}_{k}{\bm{h}}_{k}.\label{ekdef}
\end{align}
Now equation \eqref{qurt} can be used to calculate
$\mathrm{E}\bigl[ {{\bigl| \hat{\bm{g}}_{k}^{ls}\hat{\bm{g}}_{k}^{m} \bigr|}^{2}} \bigr]$
by substituting $\bm{x}=\hat{\bm{g}}_{k}^{ls}$,
$\bm{m}=\bm{h}_k$, $\boldsymbol{\Sigma}=\bm{S}_k$, 
$\bm{A}=\bm{D}=\bm{I}$, $\bm{B}=\bm{C}={\bm{F}}_{k}$,
$\bm{a}=\bm{d}=\bm{0}$ and $\bm{b}=\bm{c}={\bm{f}}_{k}$.
Applying the substitution to the right side of \eqref{qurt} leads to a sum of four complicated terms which are simplified as follows:
\begin{itemize}[labelindent=\parindent,leftmargin=*]
\item First, the expression ${\bm{A}}^{H}\bm{B}\boldsymbol{\Sigma}{{\bm{C}}^{H}}\bm{D}\boldsymbol{\Sigma}$ which equals ${\bm{F}}_{k}\bm{S}_k{\bm{F}}_{k}^{H}\bm{S}_k$ is simplified as $\bm{R}_{k}\bm{S}_k^{-1}\bm{R}_{k}\bm{S}_k$.
This is the result of applying \eqref{dkdef} and considering that ${\bm{R}}_{k}$ and 
${\bm{S}}_{k}$ are self-adjoint matrices.
The product $\bm{S}_k^{-1}\bm{R}_{k}\bm{S}_k$ is simplified as
$(\bm{R}_{k}^{-1}\bm{S}_k)^{-1}\bm{S}_k=\bm{R}_{k}$.
It follows that ${\bm{A}}^{H}\bm{B}\boldsymbol{\Sigma}{{\bm{C}}^{H}}\bm{D}\boldsymbol{\Sigma}=\bm{R}_{k}^2$ in this case.
\item 
The second expression ${{\bigl( \bm{Cm}+\bm{c} \bigr)}^{H}}\bm{D}\boldsymbol{\Sigma}{{\bm{A}}^{H}}\bigl( \bm{Bm}+\bm{b} \bigr)$ is equal to
${\bigl( {\bm{F}}_{k}{\bm{h}}_{k}+{\bm{f}}_{k} \bigr)}^{H}
{\bm{S}}_{k}
\bigl( {\bm{F}}_{k}{\bm{h}}_{k}+{\bm{f}}_{k} \bigr)$.
We conclude from \eqref{ekdef} that ${\bm{F}}_{k}{\bm{h}}_{k}+{{\bm{f}}_{k}}={\bm{m}}_{k}$.
It follows that ${{\bigl( \bm{Cm}+\bm{c} \bigr)}^{H}}\bm{D}\boldsymbol{\Sigma}{{\bm{A}}^{H}}\bigl( \bm{Bm}+\bm{b} \bigr)={\bm{m}}_{k}^{H}
{\bm{S}}_{k}{\bm{m}}_{k}$.
\item 
For the third expression, we have:
${{\bigl( \bm{Am}+\bm{a} \bigr)}^{H}}\bm{B}\boldsymbol{\Sigma}{{\bm{C}}^{H}}\bigl( \bm{Dm}+\bm{d} \bigr)={\bm{h}}_{k}^{H}{\bm{F}}_{k}\bm{S}_k{\bm{F}}_{k}{\bm{h}}_{k}$.
The product ${\bm{F}}_{k}\bm{S}_k{\bm{F}}_{k}$ is equal to $\bm{R}_k\bm{S}_{k}^{-1}{\bm{R}}_{k}$.
According to \eqref{meanmmse}, it equals ${\bm{U}}_{k}$.
Consequently, it is concluded that
${{\bigl( \bm{Am}+\bm{a} \bigr)}^{H}}\bm{B}\boldsymbol{\Sigma}{{\bm{C}}^{H}}\bigl( \bm{Dm}+\bm{d} \bigr)={\bm{h}}_{k}^{H}{\bm{U}}_{k}{\bm{h}}_{k}$.
\item 
The fourth expression is a product of two terms that each contains three parts.
From previous expression derivation:
$\bm{Bm}+\bm{b}=\bm{Cm}+\bm{c}={\bm{m}}_{k}$,
$\bm{Am}+\bm{a}=\bm{Dm}+\bm{d}={\bm{h}}_{k}$ and
$\bm{B}\boldsymbol{\Sigma}{{\bm{A}}^{H}}=
\bigl(\bm{D}\boldsymbol{\Sigma}{{\bm{C}}^{H}}\bigr)^H$.
Hence, we have:
$\mathrm{Tr}\bigl\{\bm{B}\boldsymbol{\Sigma}{{\bm{A}}^{H}}\bigr\}=
\mathrm{Tr}\bigl\{\bm{D}\boldsymbol{\Sigma}{{\bm{C}}^{H}}\bigr\}^{*}$ (The notation $^*$ denotes complex conjugate).
Therefore, the fourth expression is a product of a term with its complex conjugate.
We also have: $\bm{B}\boldsymbol{\Sigma}{{\bm{A}}^{H}}={\bm{F}}_{k}{\bm{S}}_{k}={\bm{R}}_{k}$.
Consequently, the fourth expression in \eqref{qurt} is
${\big|\mathrm{Tr}\bigl\{ {{\bm{R}}_{k}} \bigr\}
    +{\bm{h}}_{k}^H{\bm{m}}_{k}\big|}^{2}$.
\end{itemize}
Finally, inserting all terms in \eqref{qurt}, gives \eqref{numofls}.
\subsection{Appendix 4: Equation \eqref{ea1B1} proof}\label{proofmseapp}
First, $\mathrm{E}\bigl[ {\bigl| {{\bigl( \hat{\bm{g}}_{k}^{ls} \bigr)}^{H}}\hat{\bm{g}}_{i}^{m} \bigr|}^{2}\bigr]$ is derived for specific $i$ and $k$ such that $i \neq k$.
For this purpose, the inside expression is modified.
The expression ${\bigl| {{\bigl( \hat{\bm{g}}_{k}^{ls} \bigr)}^{H}}\hat{\bm{g}}_{i}^{m} \bigr|}^{2}$ is scalar and equals its trace.
By using the commutative property of trace we have
\begin{equation}\label{insidetr}
    {\left| {{\left( \hat{\bm{g}}_{k}^{ls} \right)}^{H}}\hat{\bm{g}}_{i}^{m} \right|}^{2}
    =\mathrm{Tr}\left\{ \hat{\bm{g}}_{k}^{ls}{{\left( \hat{\bm{g}}_{k}^{ls} \right)}^{H}}\hat{\bm{g}}_{i}^{m}{{\left( \hat{\bm{g}}_{i}^{m} \right)}^{H}} \right\}.
\end{equation}
Since expectation and trace are both linear operators, changing their order does not change the result.
Hence, we apply expectation to expression inside of trace \eqref{insidetr}.
By using independence of $\hat{\bm{g}}_{i}^{m}$, $\hat{\bm{g}}_{k}^{ls}$, and also \eqref{quadmat}, we have
\begin{equation}\label{expinsidetr}
    \mathrm{E}\left[{\left| {{\left( \hat{\bm{g}}_{k}^{ls} \right)}^{H}}\hat{\bm{g}}_{i}^{m} \right|}^{2} \right]=
    \mathrm{Tr}\left\lbrace \left( {{\bm{S}}_{k}}+{{\bm{h}}_{k}}\bm{h}_{k}^{H} \right)\left( {{\bm{U}}_{i}}+{{\bm{m}}_{i}}\bm{m}_{i}^{H} \right)\right\rbrace.
\end{equation}
Finally, the equation \eqref{ea1B1} is derived as follows
\begin{align}
    I_1^{ls}&=\sum\limits_{i \neq k}p_{li}\mathrm{E}\left[ {\left| {{\left( \hat{\bm{g}}_{k}^{ls} \right)}^{H}}\hat{\bm{g}}_{i}^{m} \right|}^{2} \right]\nonumber\\
    & =\sum\limits_{i \neq k}p_{li}\mathrm{Tr}\left\{ \left( {{\bm{S}}_{k}}+{{\bm{h}}_{k}}\bm{h}_{k}^{H} \right)\left( {{\bm{U}}_{i}}+{{\bm{m}}_{i}}\bm{m}_{i}^{H} \right) \right\}.
\end{align}
\subsection{Appendix 5: Off Diagonal entries effect}\label{offdiag}
In the proposed approximation, three different forms of expressions are observed as below:
\begin{align}
\label{trA2}\mathrm{Tr}\left\{ {{\bm{A}}^{2}} \right\}&=\sum_{i=1}^{M}a_{ii}^2+2
\sum_{i=1}^{M}\sum_{j=1}^{i-1}{\left|a_{ij}\right|}^2,\\
\label{xHAx}
{\bm{x}}^H\bm{Ax}&=\sum_{i=1}^{M}a_{ii}{\left|x_{i}\right|}^2+2\Re
\left(\sum_{i=1}^{M}\sum_{j=1}^{i-1}a_{ij}x_i^*x_j\right),\\
\label{trAB}\mathrm{Tr}\left\{ \bm{AB} \right\}&=\sum_{i=1}^{M}a_{ii}b_{ii}+2\Re
\left(\sum_{i=1}^{M}\sum_{j=1}^{i-1}a_{ij}^*b_{ij}\right),
\end{align}
where $\bm{A},\bm{B}$ are hermitian matrices and
$\bm{x}$ is a vector.
The entries $a_{ij}$ and $b_{ij}$ are $(i,j)$th element of $\bm{A}$ and $\bm{B}$, respectively.
${x}_{i}$ is the $i$th element of vector $\bm{x}$.
The notation $\Re (\cdot)$ denotes the real part of the inside argument.
In the above expressions, the first term on the right side is equal for both diagonal and non-diagonal matrices,
while the second term is different.
In \eqref{trA2}, the second term is always positive for a non-diagonal matrix.
Hence, $\mathrm{Tr}\big\{ {{\bm{A}}^{2}} \big\}$ is greater for a non-diagonal matrix.
However in \eqref{xHAx} and \eqref{trAB}, the second term  is not always positive.
Therefore the values of these expressions for diagonal and non-diagonal matrices depend on the second term sign.
If this term is positive, then  ${\bm{x}}^H\bm{Ax}$ is greater for a non-diagonal matrix, as well as $\mathrm{Tr}\big\{ \bm{AB} \big\}$.
If this term is negative, it is vice versa.
\end{document}